\documentclass[]{pasj01}
\usepackage{color}
\usepackage{bm}
\usepackage{lineno}

\begin{document} 
\Received{yyyy/mm/dd}
\Accepted{yyyy/mm/dd}

\title{{Large-scale magnetic field structure of NGC 3627 based on magnetic vector map}}

\author{Kohei \textsc{Kurahara}\altaffilmark{1}%
}
\altaffiltext{1}{Graduate School of Science and Engineering, Kagoshima University, 1-21-35 Korimoto, Kagoshima 890-0065, Japan}
\email{k8791902@kadai.jp}

\author{Hiroyuki \textsc{Nakanishi}\altaffilmark{1}}

\author{Yuki \textsc{Kudoh}\altaffilmark{1}}



\maketitle

\begin{abstract}
We analyzed the data of Stokes $I$, $Q$, and $U$ in C- and X-bands and investigated the large-scale magnetic field structure of NGC 3627. The polarization intensity and angle in each band were derived using Stokes $Q$ and $U$ maps. The rotation measure was calculated using the polarization-angle maps. Moreover, the magnetic field strength was calculated by assuming energy equipartition with the cosmic ray electrons. The structure of the magnetic field was well aligned with the spiral arms, which were consistent with those in the former studies. We applied the magnetic vector reconstruction method to NGC 3627 to derive a magnetic vector map, which showed that northern and southern disks were dominant with inward and outward magnetic vectors, respectively. Furthermore, we discussed the large-scale structure of the magnetic field in NGC 3627 and observed that the structure is bi-symmetric spiral in nature, and that the number of magnetic field mode is $ m_{\rm B} = 1 $ in outer region of galaxy. In addition, NGC 3627 has a mode of two spiral arms that were clearly visible in an optical image. The ratio of the mode of spiral arms to that of magnetic field is 2:1. In terms of NGC 3627, the large-scale magnetic field may be generated via the parametric resonance induced by the gravitational potential of the spiral arms.
\end{abstract}

\section{Introduction}
Although we know that magnetic fields ubiquitously exist in the universe, from the stellar scale to the large-scale structure, the understanding of the role of the magnetic field is not yet complete \citep{2016A&ARv..24....4B}. Large-scale magnetic fields of nearby galaxies are well aligned with the galactic spiral arms, which can be well traced in the optical or infrared image \citep{1982A&A...106..121B,1999A&A...348..405H,2015A&A...578A..93B}. The origin of the regular magnetic field is not understood, whether it is the amplified primordial magnetic field or that generated by a mean-field dynamo \citep{2017ARA&A..55..111H}. \par
The structure of magnetic fields is studied on the basis of the observations of synchrotron radiation. Notably, this radiation has a polarized component, and the orientation of the magnetic field can be derived by rotating the polarization angle by $ 90^\circ $. The sign of rotation measure ($RM$) indicates the direction of the line-of-sight component of the magnetic field. Positive and negative $RM$s indicate that the magnetic field is directed toward us and away from us, respectively. The absolute value of $RM$ is represented by the product of the number density of thermal electrons and the magnetic field strength along the line-of-sight. There exists a method of examining the magnetic field structure of nearby galaxies by using the azimuthal variation in $RM$ \citep{1985A&A...144..257S}. Accordingly, the regular magnetic field structure of some nearby galaxies was studied using the method \citep{2013pss5.book..641B}. A magnetic field structure is classified as axis-symmetric spiral (ASS), which has a pitch angle and only either an outward or inward magnetic component, or bi-symmetric spiral (BSS), which has a pitch angle and both outward and inward magnetic components. Some studies showed magnetic field vectors without the $180^\circ$ ambiguity by comparing multiple frequency data with theoretical models \citep{2004A&A...414...53F,2005A&A...444..739B}. Reportedly, the M51 disk and halo structures were studied in the same method \citep{2011MNRAS.412.2396F}. Recently, a magnetic vector reconstruction method was developed by \citet{2019Galax...7...32N}. Notably, \citet{2019Galax...7...59K} applied the method to NGC 6946 to study the large-scale magnetic field thereof.\par
In this study, we focus on NGC 3627, which is located at (RA (J2000), Dec (J2000)) = (11h20m15.027s, +12d59m29.58s) and is classified as a molophological type of SAB(s)b \citep{1991rc3..book.....D}. In addition, its galaxy distance is 11.1 Mpc \citep{2019ApJ...882..150H}, at which the angular size of 1" is equivalent to the physical scale of 54 pc. The position angle and inclination of NGC 3627 are $176^\circ$ and $52^\circ$, respectively \citep{2007PASJ...59..117K}. These basic pieces of information regarding NGC 3627 are listed in table \ref{tab:1}. NGC 3627 is one of the Leo Triplet group galaxies \citep{1966ApJS...14....1A}. Its stellar disk has been reported to have two strong spiral arms \citep{2014ApJ...790..118M}.\par
The polarized synchrotron emission of this source has been used to study its magnetic field by \citet{1999A&A...345..461S}, who reported the total intensity and polarization maps of NGC 3627 on the basis of the X-band observation using the Effelsberg 100-m telescope. The polarization intensity was smoothly distributed in the galactic disk, and strong polarization component in the dust lane of the western arm and weak component between the spiral arms in the northeast. Additionally, on the basis of the C- and X-band observations, \citet{2001A&A...378...40S} reported that the direction of the magnetic field was well aligned with the arm on the western side of the disk. \par 
In this paper, we report on the magnetic field structure of NGC 3627; the structure was obtained by applying the magnetic vector reconstruction method to NGC 3627. In Section 2, we summarize the data used in this study, and in Section 3, we describe the magnetic vector reconstruction method. Furthermore, in Section 4, we show the results, which include the magnetic field strength and magnetic vector map. In Section 5, we discuss the results obtained. Finally, in Section 6, we provide a summary.\par

\begin{table}[h]
  \tbl{Basic parameters of NGC 3627}{%
  \begin{tabular}{lll} \hline
Parameter & Value & reference \\ \hline
RA (J2000)& $11^h20^m15.027^s$ & \citet{1991rc3..book.....D} \\ 
Dec (J2000)& $+12^d59^m29.58^s$ & \citet{1991rc3..book.....D} \\
Morphology & SAB(s)b & \citet{1991rc3..book.....D} \\
Distance & 11.1 Mpc & \citet{2019ApJ...882..150H} \\
Position Angle& $176^\circ$ & \citet{2007PASJ...59..117K} \\
Inclination & $52^\circ$ & \citet{2007PASJ...59..117K} \\ \hline
    \end{tabular}}\label{tab:1}
\end{table}

\section{Data}
\begin{figure*}[h]
 \begin{center}
  \includegraphics[width=8.3cm]{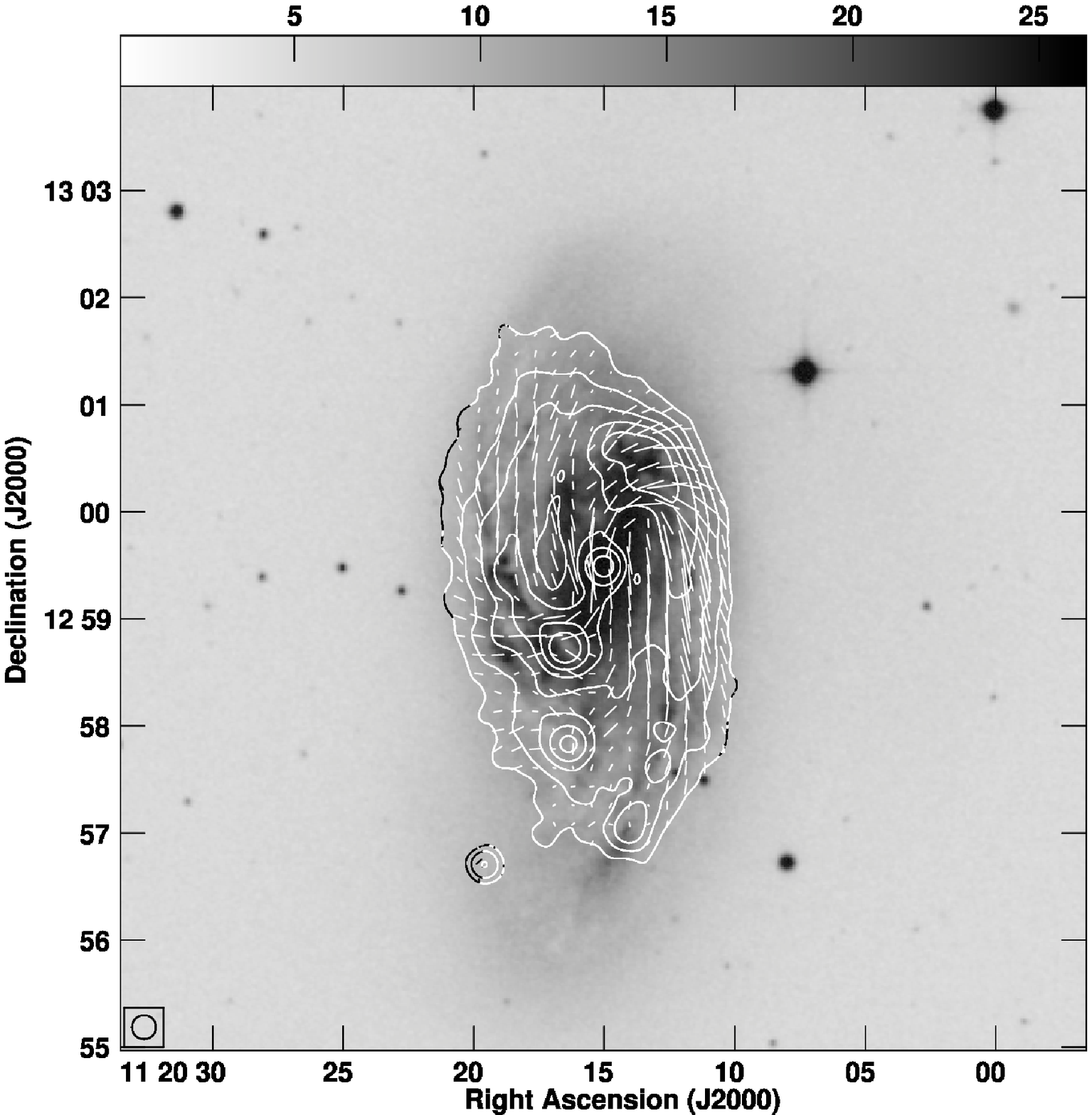} 
  \includegraphics[width=8.3cm]{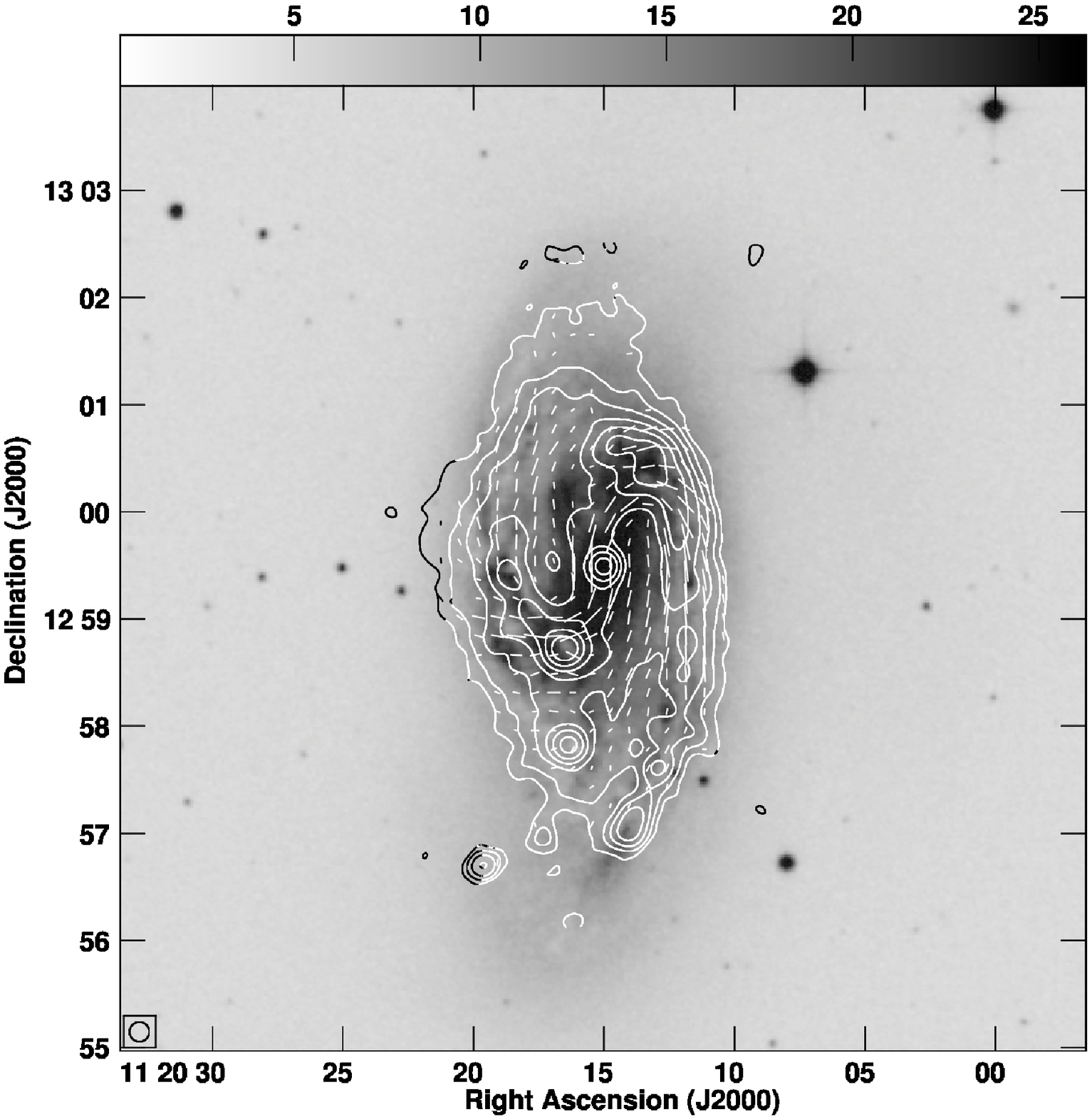} 
 \end{center}
\caption{Maps of the total intensity (contour) and orientation of the magnetic field (white bars) of the C- (left) and X- (right) bands of NGC 3627. The background is a black-and-white image of DSS2B \citep{2000ASPC..216..145M}. The orientation of magnetic field obtained by rotating the observed polarization plane by $ 90^\circ $. The white lines are plotted every 7 arcsec in the C- band and 5 arcsec in the X-band, by considering the spatial resolution. A beam is drawn at the lower left of each map, and beam sizes are 13.5 and 11.0 arcsec, respectively. The contour intervals are (3, 6, 12, 24, 48, 96) $ \times$ 66 $ \mu $Jy / beam, (3, 6, 12, 24, 48, 96) $ \times $ 27 $ \mu $Jy / beam.}\label{fig:radio image}
\end{figure*}

\begin{figure}[h]
 \begin{center}
  \includegraphics[width=8.5cm]{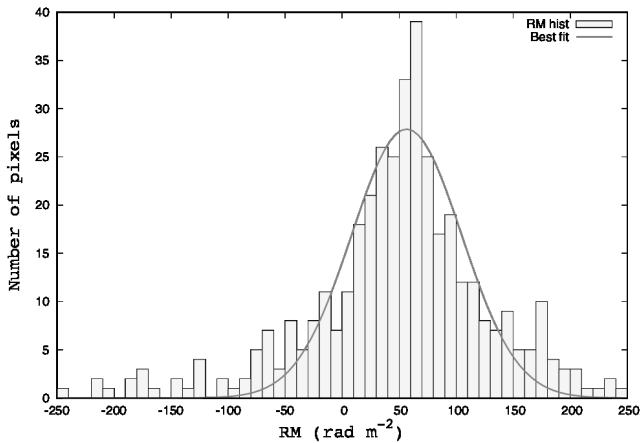} 
 \end{center}
\caption{$RM$ histogram of NGC 3627. The horizontal axis represents $RM$ $ {\rm rad\ m^{-2}} $ and the vertical axis represents the number of pixels. The Gaussian that was the best fit within the absolute value of RM within $250\ {\rm rad\ m^{-2}} $ was drawn using a solid line. The horizontal axis is drawn with a resolution of $ 10 {\rm rad\ m^{-2}} $.}\label{fig:RM histogram}
\end{figure}

\begin{figure}[h]
 \begin{center}
  \includegraphics[width=8cm]{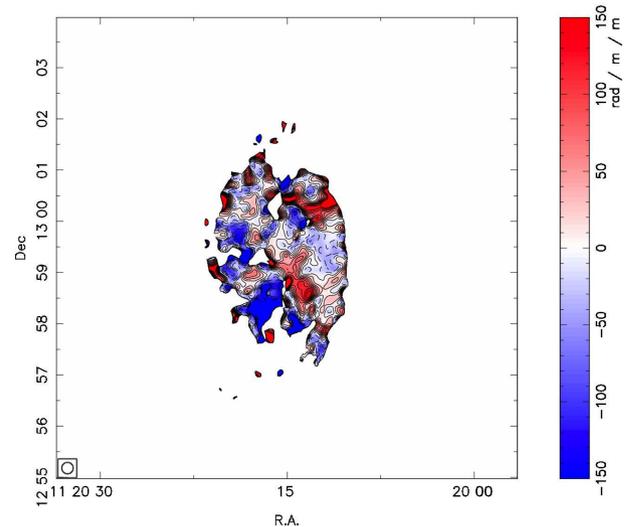} 
 \end{center}
\caption{$RM$ map of NGC 3627 at angular resolution 13.5 arcsec. $RM$ $ + 56.5 {\rm rad\ m^{-2}} $ in the foreground was removed. The pixels that had the absolute value of $RM$ greater than 200 were masked as unrealistic.}\label{fig:RM map}
\end{figure}

\subsection{Data reduction}
The data used in this study were published by \citet{2001A&A...378...40S} and obtained from observations in the C-band (4.85 GHz) and X-band (8.46 GHz). They consist of a combination of single dish and interferometry data, obtained with the Effelsberg radio telescope and very large array (VLA), respectively. Observational information, such as the frequency, size of the synthesized beam, and root mean square (rms) noise, are summarized in Table \ref{tab:data parameter}.\par

\begin{table}[h]
  \tbl{Information of the data and data errors.}{
  \begin{tabular}{lcc} \hline 
& C & X \\ \hline
Frequency & 4.85 GHz &8.46 GHz \\
Beam size & 13.5" $\times$ 13.5" & 11.0" $\times$ 11.0" \\
Stokes $I$ rms & 66 $\rm{ \mu Jy\ beam^{-1}}$ & 27 $\rm{ \mu Jy\ beam^{-1}}$ \\
Stokes $Q$ rms &11 $\rm{ \mu Jy\ beam^{-1}}$ & 6 $\rm{ \mu Jy\ beam^{-1}}$ \\
Stokes $U$ rms &10 $\rm{ \mu Jy\ beam^{-1}}$ & 6 $\rm{ \mu Jy\ beam^{-1}}$ \\ \hline
\end{tabular}}\label{tab:data parameter}
\end{table}

Data analysis was performed using the software AIPS (Astronomical Image Processing System). We calculated the polarization intensity and polarization angle using the AIPS task COMB. Using the Stokes parameter, the polarization intensity and angle were calculated to be $ P_{\rm POL} = \sqrt {Q^2 + U^2} $ and $ \chi = \frac{1}{2}{\rm arctan} ( U / Q ) $, respectively. The polarization intensity was corrected by subtracting the positive bias using the task POLCO. The primary beam was also corrected using the task PBCORL. The noise level $\sigma$ of each Stokes parameter was the variance of the emission-free region and measured using TVBOX and IMSTAT, both of which are the verbs of AIPS. The measured noise levels are summarized in table \ref{tab:data parameter}. The detection limit was defined as thrice the noise level, i.e., $3 \sigma$.

\subsection{Radio images}
We derived the total and polarized intensities in the C- and X-bands. In figure \ref{fig:radio image}, we depict the contours of the total intensities and orientations of B-field ($90^\circ$ rotation of the E-field) overlaid on a black-and-white image from DSS2 B \citep{2000ASPC..216..145M}. The observed B-field is plotted only in pixels which the polarized intensities were detected with higher signal-to-noise (S/N) ratio than the detection limit in Stokes $Q$ and $U$ ($> 3(\sigma_{\rm Q} + \sigma_{\rm U})/2$).\par
The total intensity is the brightest in the boundary region between the southern bar and spiral arm, and its values are 12.99 and 9.19 mJy/beam at the C- and X-bands, respectively. This region is also bright in CO and HI lines \citep{2007PASJ...59..117K, 2008AJ....136.2563W}. The polarized intensity is brightest in the galactic central region in the X-band and in the south-western spiral arm in the C-band.

\subsection{Rotation measure}
To calculate the rotation measure ($RM$), Stokes $I$, $Q$ and $U$ maps of the X-band with higher angular resolution were convolved into a 13.5" $\times$ 13.5" beam using the AIPS task CONVL. We calculated the $RM$ at the pixels, which the polarized intensity was detected with better S/N ratio than the detection limit in both bands, based on the following equation:

\begin{equation}
\frac{RM}{\rm rad \, m^{-2}} = \frac{\chi (\lambda _1) - \chi (\lambda _2)}{\lambda _1 ^2 - \lambda _2 ^2} ,
\label{equ:RM}
\end{equation}

\noindent where $\chi (\lambda)$ denotes the polarization angle at wave length $\lambda$.\par
The histogram of the calculated $RM$ is depicted in figure \ref{fig:RM histogram}. We fitted the histogram of $RM$ with a Gaussian function by using the nonlinear least squares method, which is based on the Levenberg--Marquardt method. The peak value of the fitted Gaussian was 27.9 at $+ 56.5 \ {\rm rad\ m^{-2}}$, and the dispersion was $48.3 \ {\rm rad\ m^{-2}}$, which is consistent with the value of galaxies \citep{2016A&ARv..24....4B}. We considered that $+ 56.5 \ {\rm rad\ m^{-2}}$ was attributed to the foreground.\par 
In figure \ref{fig:RM map}, we depict an $RM$ map of NGC 3627, which the foreground $RM$ of $+ 56.5 \ {\rm rad\ m^{-2}}$ was subtracted. The color range indicates the $RM$ value. The contour levels are $-$150.0, $-$131.25, $-$112.5, $-$93.75, $-$75.0, $-$56.25, $-$37.5, $-$18.75, 0.0, 18.75, 37.5, 56.25, 75.0, 93.75, 112.5, 131.25, and 150.0 ${\rm rad\ m^{-2}}$. This distribution is consistent with that reported by \citet{2001A&A...378...40S}.

\begin{figure*}[hp]
 \begin{center}
  \includegraphics[width=14cm]{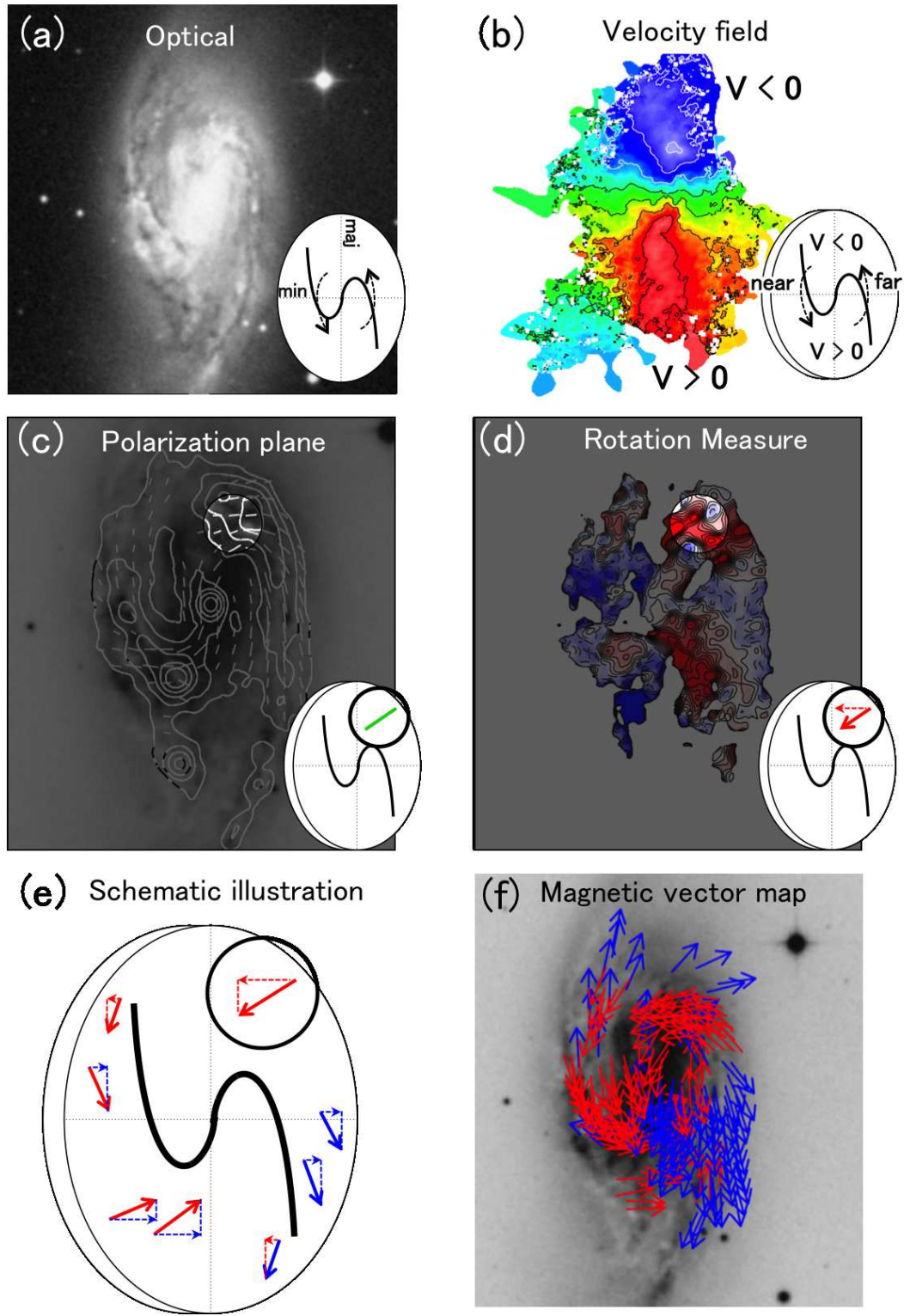} 
 \end{center}
\caption{Steps in the magnetic field vector reconstruction method \citep{2019Galax...7...32N}. (a) Determine the direction of rotation of the galaxy on the basis of an optical image after assuming a trailing spiral. (b) Determine the near side of the galaxy by combining the map of the velocity field and the direction of rotation. (c) Determine the direction of the magnetic field perpendicular to the line-of-sight from the polarization plane of the synchrotron (having an ambiguity of $ 180^\circ $ along the line of sight). (d) Solve the ambiguity of $ 180^\circ $ along the line-of-sight using the sign of $RM$.(e) Determine the direction of all magnetic vectors. (f) A magnetic vector map of NGC 3627 can be derived.}\label{fig:How_to_decide_the_vector}
\end{figure*}

\section{Magnetic vector reconstruction method}
We used the magnetic vector reconstruction method, which is described in \citet{2019Galax...7...32N}. The method can be used to determine the directions of magnetic vectors without the $180^\circ$ ambiguity by using observational data. The directions of magnetic vectors are derived from the multifrequency radio data of synchrotron polarization after determining the orientation of a disk on the basis of the shape of spiral arms and velocity field. The advantage of this method is that the magnetic field vector of a galaxy can be determined using only simple assumptions and observational data. More details are described as follows.\par
First, we must examine the direction of galactic rotation. Because spiral galaxies generally have trailing spiral arms \citep{2019ApJ...886..133I}, NGC 3627 rotates counter-clockwise in the sky plane, as depicted in figure \ref{fig:How_to_decide_the_vector} (a). Next, we used the velocity map of HI data for determining the geometrical orientation. We observed that the east side was the near side, as the northern and southern parts were blue- and red-shifted, respectively, as depicted in figure \ref{fig:How_to_decide_the_vector} (b). The orientations of the magnetic field at individual pixels are given by the synchrotron polarization angle of a higher-frequency band, with the $180^\circ$ ambiguity, as depicted in figure \ref{fig:How_to_decide_the_vector} (c). This ambiguity was solved by checking the sign of RM, as depicted in figure \ref{fig:How_to_decide_the_vector} (d). If the RM of a pixel is positive, the magnetic-field component along the line-of-sight of the pixel is directed toward us from the source, and vice versa. Because the eastern side of the disk is close to us, as previously mentioned, the vector is directed south-eastward, as depicted in figure 4 (d). Therefore, the directions of all the magnetic vectors can be determined point-by-point, as depicted in figure 4 (e), where the inward and outward vectors were plotted in red and blue, respectively. Finally, we can derive a magnetic vector map, as depicted in figure 4 (f).\par
Notably, in our analysis, we ignored the south-eastern region, where the HI velocity field is significantly deviated from its galactic rotation. This region is not detected in synchrotron emission.\par

\section{Magnetic field of NGC 3627}
\subsection{Strength of magnetic field}
We derived the strength of the total and ordered components of the magnetic field by using the total intensity and its polarization degree, respectively. The total magnetic field strength ($ B_{\rm tot} $) is calculated using equation (\ref{equ:strength}) based on an assumption of energy equipartition between cosmic rays and its magnetic field described in \citet{2005AN....326..414B},

\begin{equation}
B_{\rm tot} = \left\{ \frac{4 \pi (2\alpha + 1)( K_0 + 1)I_\nu E_{\rm p}^{1-2\alpha}(\nu/2c_1)^\alpha}{(2\alpha -1)c_2(\alpha)L\ c_4(i)} \right\} ^{1/(\alpha+3)} ,
\label{equ:strength}
\end{equation}

\noindent where $\alpha$ denotes the spectral index of synchrotron radiation, $K_0$ the number densities ratio of cosmic ray nuclei to that of the electrons, $L$ the path length of the synchrotron emitting media, $I_\nu $ the intensities at frequency $ \nu $, and $E_{\rm p} $ the proton rest energy, respectively. In addition, coefficients $ c_2 $ and $ c_4(i)$ denote the spectral index and galaxy inclination, and $c_1 = 3e/(4\pi m_{\rm e}^3c^5) = 6.3 \times 10^{18}\ {\rm erg^{-2} s^{-1} G^{-1}}$.\par
The strength of the ordered component can be obtained by using the equation $B_{\rm ord} = B_{\rm tot} / (1+q^2)$, where $q$ is the ratio of the turbulent to ordered components. The ratio $q$ is given by $ q \simeq \sqrt{2.1\ p_0/p} $, where $p$ is observed polarization degree and $p_0$ is intrinsic polarization degree. The polarization degree $p$ was calculated using Stokes $I$ and $P_{\rm POL,corrected} $ values ($ p = P_{\rm POL,corrected} / I$). $P_{\rm POL,corrected}$ is corrected polarization intensity of positive bias. The intrinsic polarization degree $ p_0 $ is determined using the spectral index as $p_0=(3-3\alpha)/(5-3\alpha)$ , as described in \citet{2005AN....326..414B}. We adopted $\alpha = 0.9$, $K_0=100$, $L=6.2\times 10^{21}\ {\rm cm}$, $E_{\rm p} = 1.5 \times 10^{-3}\ {\rm erg}$, $c_2 = 3.9 \times 10^{-24}\ {\rm erg~G^{-1} sterad^{-1}}$, and $c_4 = 0.62^2$, to estimate the magnetic field strength, by following former works \citep{2001A&A...378...40S, 2005AN....326..414B}. \par
\begin{figure}[h]
 \begin{center}
  \includegraphics[width=8cm]{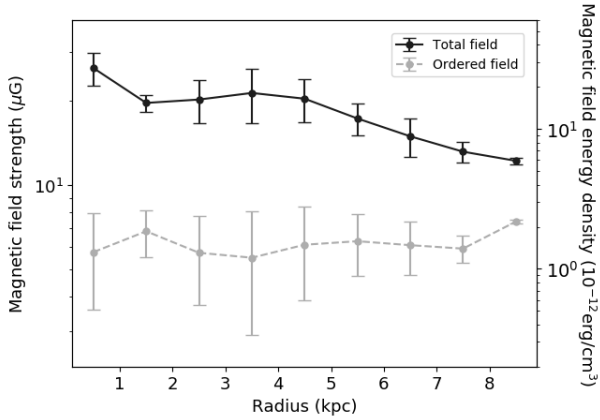} 
 \end{center}
\caption{Radial distribution of the magnetic field strength. The horizontal axis represents the radius from the galactic center, in kpc, and the left and right sides of the vertical axis represent the magnetic field strength in $\mu$G and the energy density equivalent to the magnetic field strength of $10^{-12}$erg\ cm$^{-3}$, respectively. The black solid line indicates the total magnetic field strength and the grey dashed line indicates the ordered field.}\label{fig: magnetic field strength}
\end{figure}

The radial distribution of the magnetic field strength and its equivalent energy density is depicted in figure \ref{fig: magnetic field strength}. The averaged magnetic field strength over the galaxy is $ B_{\rm tot} = 18.8 \pm 4.2\ \mu$G for the total field, and $ B_ {\rm ord} = 6.1 \pm 2.0\ \mu$G for the ordered field. Although the total field is the highest at the galactic center and decreases with radius, it is roughly constant in the range of r = 1.5--5.5 kpc. The ordered field is almost constant for the all the radii. Error bars are the standard deviation of each data within the corresponding radius range.

\citet{2001A&A...378...40S} observed that $B_{\rm tot} = 11 \pm 2\ \mu$G for the total field, and $B_{\rm ord} = 4 \pm 1\ \mu$G for the ordered field, both of which are different than our results. However, we also observed that $B_{\rm tot, class} =13.2 \pm 3.3\ \mu$G and $B_{\rm ord, class} = 7.3 \pm 1.2\ \mu$G using the classical method described by \citet{2005AN....326..414B},

\begin{equation}
B_{\rm tot, class} = \left\{ \frac{6 \pi (\kappa + 1)I_\nu (\nu /2)^\alpha c_1^{-1/2} \left[ \nu _{\rm min}^{1/2-\alpha } - \nu_{\rm max}^{1/2-\alpha} \right] }{(2\alpha -1)c_2(\alpha)L\ c_4(i)} \right\} ^{2/7},
\label{equ:strength_class}
\end{equation}

\noindent where $ \nu_{\rm min} $ and $ \nu_{\rm max} $ denote the ranges of the spectrum for synchrotron radiation.\par
In \citet{2001A&A...378...40S}, the cutoff value of the lower electron energy was used to be 300 MeV. Therefore, we adapted $\nu_{\rm min} = 500$ MHz, which was converted from 300 MeV (e.g., equation 2 of Reynolds \& Keohane 1999), and $\nu_{\rm max} = 10$ GHz. Furthermore, when we calculated $ B_{\rm ord, class} $ using the classical method, the polarization intensity was used directly as $ I_\nu $ instead of dividing it by $q$. From the above-mentioned results, it is considered that the difference between our results and that of \citet {2001A&A...378...40S} is attributed to the difference between the methods used.

\subsection{Magnetic field vector map}
\begin{figure*}[h]
 \begin{center}
  \includegraphics[width=8cm]{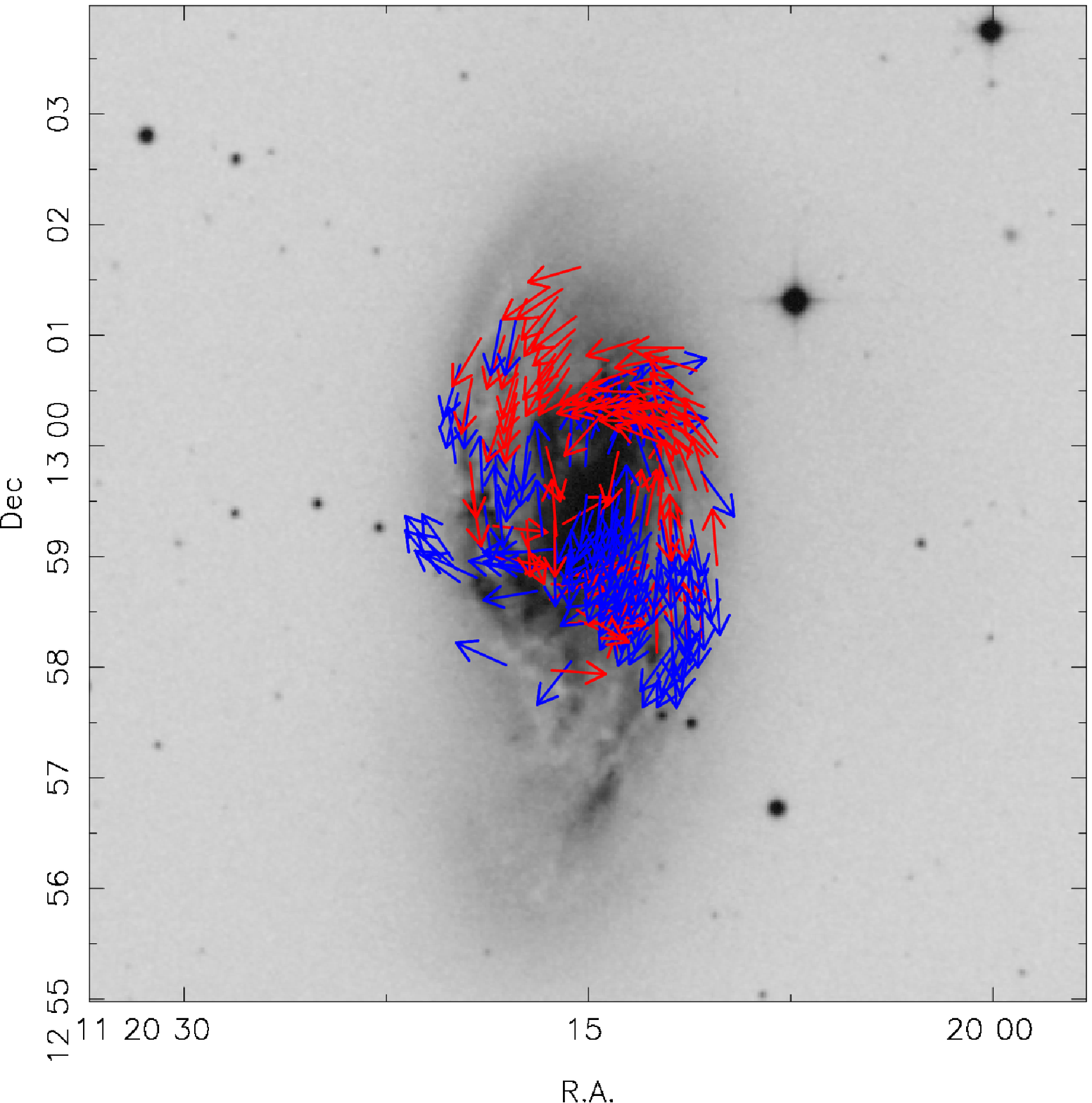} 
  \includegraphics[width=8cm]{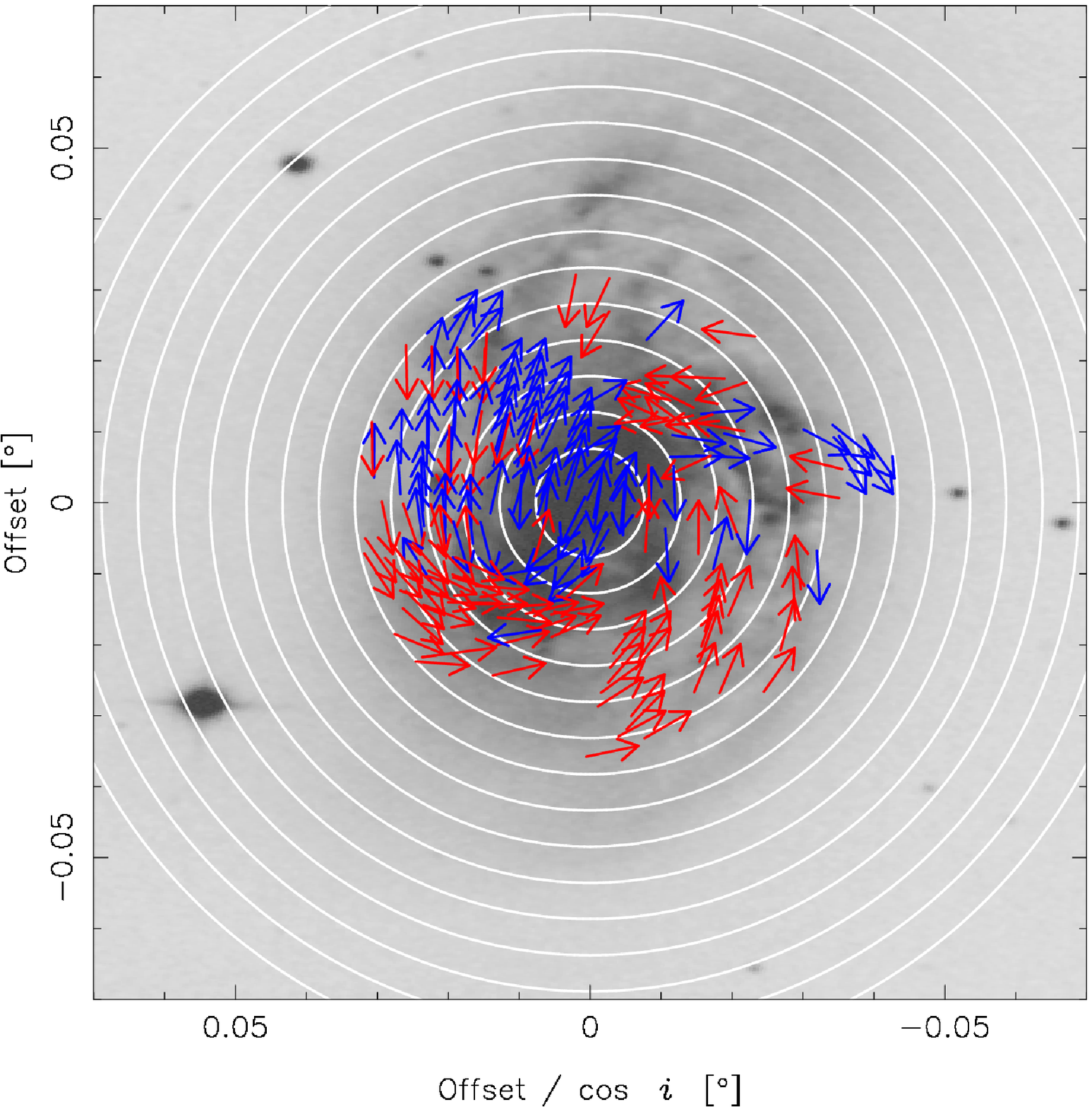} 
 \end{center}
\caption{(Left) Magnetic field vector map of NGC 3627 in the sky plane. Effective angular resolution is 13.5 arcsec. The background is the same as that in figure \ref{fig:radio image}. The red and blue arrows show inward and outward magnetic vectors, respectively. The vectors are plotted every 7 arcsec. (Right) The face-on view of the magnetic field vector map, which made by rotating counter-clockwise by the position angle and extending in the minor-axis direction by $ {\rm 1 / cos\ } i $. The white circles are drawn at the intervals of 1.0 kpc from the center of the galaxy.}\label{fig:vector map}
\end{figure*}

We derived a magnetic field vector map using the magnetic vector reconstruction method, as explained in Section 3. The left panel of figure \ref{fig:vector map} shows a magnetic field vector map of NGC 3627. The magnetic field vectors were plotted only at the points that satisfied the following four criteria: (1) Stokes $I$ and $ P_{\rm POL}$ were higher than $3\sigma_I$ and $3(\sigma_{\rm Q} + \sigma_{\rm U})/2$ in the C- and X-bands, respectively; (2) the $RM$ value was higher than the detection limit ($\sigma_{\rm RM}$), which was calculated on the basis of error propagation using the expression $ \sigma_{\rm RM} = \sqrt{(\delta \chi (\lambda_1)^2 -\delta \chi (\lambda_2)^2)/ ({\lambda_1^2}-{\lambda_2^2})^2}$; (3) the absolute value of RM is within 200 ${\rm rad\ m^{-2}}$, and (4) the difference in polarization angle in each band is larger than the sum of polarization angle errors. We used the polarization angle in the X-band as the orientation of the magnetic vector. We neglected the Faraday rotation due to the intervening media between the source and our location, as the typical $RM$ dispersion was $48.3{\rm rad\ m^{-2}}$, which corresponded to a few degrees of rotation in the X-band.\par
In the right panel of figure \ref{fig:vector map}, we show a face-on view of the magnetic field vector map. It was made by rotating the left panel counter-clockwise by the position angle of $ 176^\circ $ to align the major axis with the vertical axis and horizontally enlarging it by a factor of 1/cos $i$. We observed that NGC 3627 has both inward (red) and outward (blue) magnetic field vectors, it suggests that the structure of the magnetic vector was not axis-symmetric spiral (ASS). In the left panel of figure \ref{fig:vector map}, we show that the vectors in the northeastern and southwestern regions are inward and outward, respectively. Therefore, the mode of bi-symmetric spiral (BSS) seems dominant in this galaxy, as discussed in the following section. 

\section{Large-scale magnetic vector structure}
\subsection{Definitions of pitch angle, vector angle, and sign of inward or outward}
\begin{figure*}[h]
 \begin{center}
  \includegraphics[width=12cm]{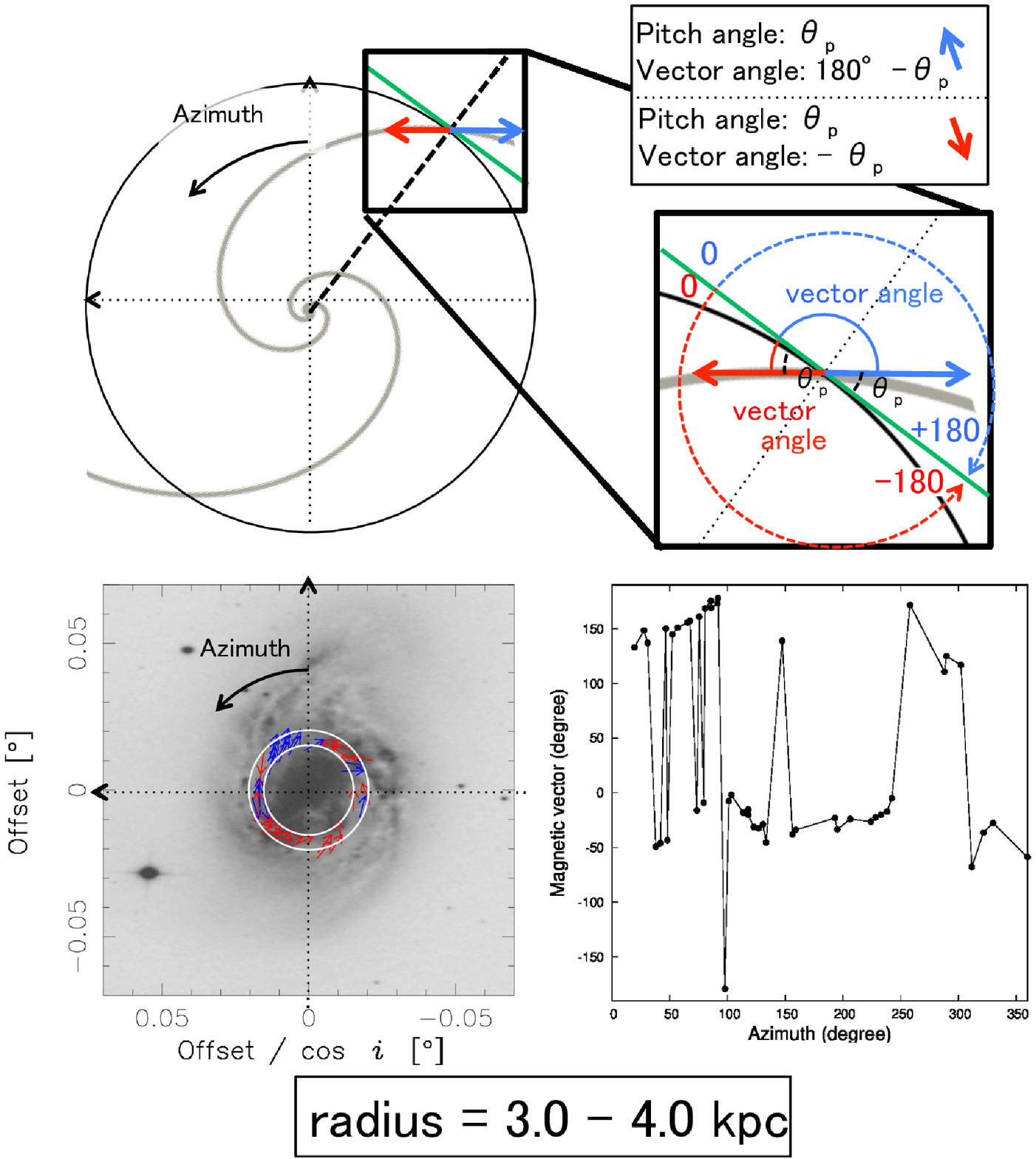} 
 \end{center}
\caption{Schematic and definition of galactic azimuth and magnetic field vector angle. Bottom-left bottom-right panels show a magnetic vector map within a ring of 3.0 -- 4.0 kpc and azimuthal profiles of magnetic vector, respectively. }\label{fig:Azimuth def}
\end{figure*}

As shown in the upper-right panel of figure \ref{fig:Azimuth def}, the angle of the magnetic vector is defined as the angle between the tangent of the circle and the magnetic vector. For instance, the pitch angle of $0^\circ$ corresponds to the vector angle of $0^\circ$ or $-180^\circ$. Azimuth is defined counter-clockwise from the major  axis of the galaxy as shown in the upper-left of figure \ref{fig:Azimuth def}. The bottom-left and bottom-right panels show the vector map within a ring of 3.0 -- 4.0 kpc, and a corresponding azimuthal profile of the magnetic vector angle, respectively.\par

\begin{figure*}[h]
 \begin{center}
  \includegraphics[width=16cm]{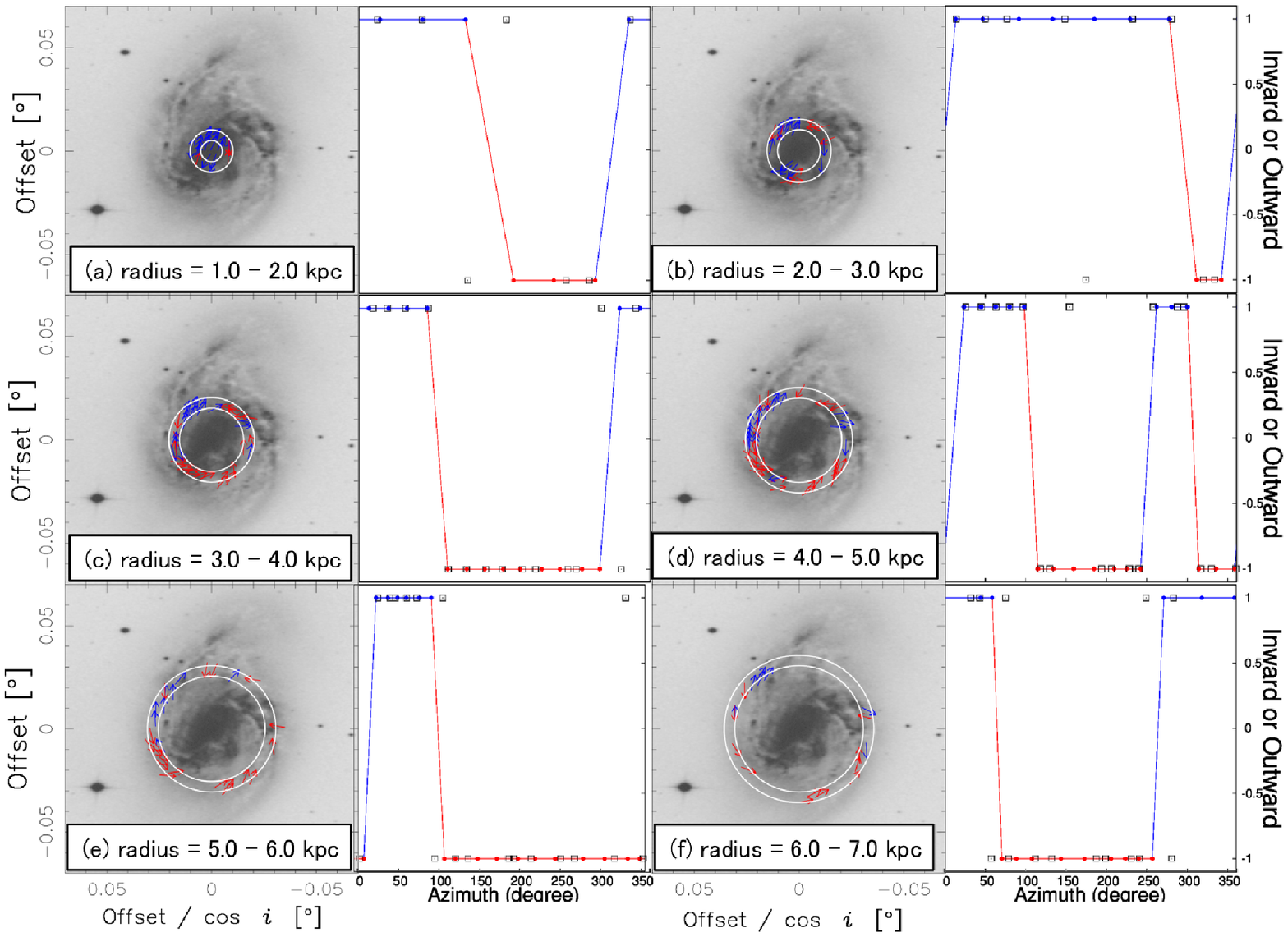} 
 \end{center}
\caption{{\bf Column 1,3:}\ Face-on views of the magnetic vector map of radius ranges of 1.0 kpc -- 7.0 kpc.\ {\bf Column 2,4:}\ Azimuthal profiles of magnetic vector angle.}\label{fig:Azimuthal_profile}
\end{figure*}

\subsection{Mode of magnetic field based on the reversal}
We studied the modes of the magnetic field by focusing on the reversals of the magnetic vector. It can be seen from figure \ref{fig:vector map} that NGC 3627 has both inward and outward magnetic vectors. Panels of the first and third columns of figure \ref{fig:vector map} show magnetic vector maps within individual rings at radial intervals $18"$ ($=1.0$ kpc), and panels of the second and fourth columns show whether magnetic vectors directed inward or outward at azimuthal resolution $13.5"$ ($=$ beam size). Inward and outward vectors in figure \ref{fig:Azimuthal_profile} were defined as $-1$ and $+1$, respectively. We considered the most frequent value of three successive data points, which was obtained by the $most\_frequent\_value( s_{j-1}, s_{j}, s_{j+1})$ when the sign of the $j$-th data point was defined as $s_j$. The red and blue dots of the azimuthal profiles indicate the inward and outward vectors, respectively.\par
If the magnetic field has mode $m_{\rm B}$, $2m_{\rm B}$ reversals can be found azimuthally. Therefore, we studied the mode of the magnetic field by counting the number of magnetic field reversals. We found $m_{\rm B}=2$ in panel (d) of figure \ref{fig:Azimuthal_profile} and $m_{\rm B}=1$ in the others. Therefore, we found that the magnetic mode of $m_{\rm B}=1$ was dominant in NGC 3627. The radius with $m_{\rm B}=2$ corresponds to the bar-end region at a radius in the range of 4.0 to 5.0 kpc (\cite{2011MNRAS.411.1409W, 2015NewA...34...65Z}). This is consistent with the result of \citet{2005A&A...444..739B}, in which regular magnetic fields change the sign in the bar-end region. For rings of radius in the range of 0.0 to 1.0 kpc and $>$ 8.0 kpc, the magnetic mode was not included because there were few data points ($< 6$ points). These results for $m_B$ are consistent with those reported by \citet{2001A&A...378...40S}, who found mixed magnetic field modes.\par
The mode of $m_{\rm B}=1$ in the outer region (r$>5.0$ kpc) corresponds to the BSS structure. NGC 3627 has two spiral arms, which means that the mode of the spiral arms is $m_{\rm D}=2$ \citep{2014ApJ...790..118M}. Therefore, we observed that the relationship between the modes of the spiral arms and the magnetic field was $m_{\rm B}=m_{\rm D}/2$ in NGC 3627. This is in good agreement with the results of analytical galactic dynamo models reported by \citet{1990MNRAS.244..714C}. They indicated that the two-armed spiral galaxies, that is, $m_{\rm D}=2$, can have the BSS structure due to the swing excitation. The growth of the magnetic field in this dynamo is described as a parametric resonance driven by the velocity profile around the density wave. It is not clear whether our results are universal or not in spiral galaxies. However, magnetic vector maps are useful for exploring reversal field structures such as BSS.\par

\subsection{Possible errors}
We understand that determination of the mode number simply by counting the number of field reversals is not a common method and that the Fourier transformation is often used. However, we noticed that the separations of spiral arms were not azimuthally equal and that the Fourier mode could be different from the actual apparent number of spiral arms. Therefore, we used this method instead of Fourier analysis to determine the relationship between the number of stellar spiral arms and the mode of the magnetic field.\par
We should also mention that the $RM$ foreground of $+ 56.5\ {\rm rad\ m^{-2}}$ estimated in our study is slightly different from \citet{2001A&A...378...40S}, who adopted $+50\ {\rm rad\ m^{-2}}$ as the $RM$ foreground. This seems attributed to difference in observational band as \citet{2001A&A...378...40S} estimated the value using L-band data taken from \citet{1980ApJ...242...74S} while we used C- and X-band data. Since L-band data are strongly affected by depolarization compared to C- and X-bands data, we conclude that our estimation of the $RM$ foreground is more suitable to our study which is based on C- and X-bands. \par
Finally, we should mention how halo component of NGC 3627 affects our results. A study by \citet{2008A&A...480...45S} showed that if the galactic inclination is more than $ 15^\circ $, the contribution of the vertical component of the halo to the observed $RM$ is negligible. Since the inclination of NGC 3627 is $ 52^\circ $, the obtained $RM$ map is not heavily affected by the halo component. In addition, \citet{2017ARA&A..55..111H} mention the observed Faraday screen may be seen as different depending on the frequency, and the Faraday screens at the C- and X-bands represented a galactic disk. In this study, because we used C- and X-bands data, we considered the contribution of the halo component to be negligible. 

\section{Summary}
We investigated the large-scale magnetic field structure of a two-armed galaxy, NGC 3627. Table \ref{tab:3} summarizes parameters of the magnetic field of NGC 3627.

\begin{table}[h]
  \tbl{Magnetic field parameters of NGC 3627}{%
  \begin{tabular}{cccc} \hline \hline
Radius kpc & $B_{\rm tot}\ \mu$G & $B_{\rm ord}\ \mu$G & $m_{\rm B}$ \\ \hline
0.0--1.0 & $26.2\pm 3.5$ & $5.6\pm 2.2$ & - \\
1.0--2.0 & $19.7\pm 1.4$ & $6.8\pm 1.3$ & 1 \\
2.0--3.0 & $20.2\pm 3.6$ & $5.7\pm 2.0$ & 1 \\
3.0--4.0 & $21.3\pm 4.7$ & $5.5\pm 2.6$ & 1 \\
4.0--5.0 & $20.4\pm 3.6$ & $6.1\pm 2.3$ & 2 \\
5.0--6.0 & $17.3\pm 2.2$ & $6.3\pm 1.8$ & 1 \\
6.0--7.0 & $14.9\pm 2.3$ & $6.1\pm 1.3$ & 1 \\
7.0--8.0 & $13.2\pm 1.1$ & $5.9\pm 0.6$ & - \\
8.0--9.0 & $12.2\pm 0.3$ & $7.4\pm 0.1$ & - \\ \hline
\end{tabular}}\label{tab:3}
\end{table}

As a result of our analysis, polarization intensity and polarization angle maps were derived from the Stokes $Q$ and $U$ maps. The spatial distribution of $RM$ was derived from the polarization angle maps. In addition, the magnetic field strength was calculated using the method by \citet{2005AN....326..414B}, and the average magnetic field strength over the map was $ 18.8\ {\rm \mu }$G and $ 6.1\ {\rm \mu }$G for the total and ordered components, respectively. We determined the direction of the galactic rotational spin using the shape and velocity field of the spiral arm by assuming the trailing arm. We derived a magnetic field vector map using $RM$ and polarization angle data on the basis of the magnetic vector reconstruction method described in \citet{2019Galax...7...32N}. \par
We observed that the pitch angle and $RM$ distributions are consistent with those in previous studies such as \citet{1999A&A...345..461S} and \citet{2001A&A...378...40S}. From the magnetic field vector map, it is observed that NGC 3627 had a $ m_{\rm B} = 1 $ mode called BSS in outer region of galaxy and inside of bar-end region. We also found that there is a locally $m_{\rm B}=2$ mode due to an additional reversal in the bar-end regions. Since NGC 3627 has two prominent spiral arms ($m_{\rm D}=2$), our result in outer region of galaxy may be consistent with theoretical model of the parametric resonance suggested by \citet{1990MNRAS.244..714C}, who predicts the magnetic field mode number ($m_{\rm B}$) is half the mode of the spiral arms ($m_{\rm D}$). \par

\begin{ack}
We are grateful to M. Soida at Astronomical Observatory of the Jagiellonian University for providing data of this study. We are grateful to M. Chiba for helpful comments. We would like to thank T. Ozawa for useful comments on our data reduction.
\end{ack}

\end{document}